\documentstyle[epsfig,prl,twocolumn,aps,floats]{revtex}
\begin{document}
\topmargin = -2.0cm
\overfullrule 0pt
\def\ltap{\ \raisebox{-.4ex}{\rlap{$\sim$}} \raisebox{.4ex}{$<$}\ }
\def\gtap{\ \raisebox{-.4ex}{\rlap{$\sim$}} \raisebox{.4ex}{$>$}\ }
\twocolumn[\hsize\textwidth\columnwidth\hsize\csname
@twocolumnfalse\endcsname
\title
{ Reply on the Comment on ``New Conditions for a Total Neutrino 
Conversion in a Medium''
}
\author
{ M. V. Chizhov$^1$ and S. T. Petcov$^{2}$}
\address
{
$^1$Faculty of Physics, University of Sofia, 1164 Sofia, Bulgaria\\
$^2$SISSA, I-34014 Trieste, Italy}
  \maketitle

\vspace{.5cm}
\hfuzz=25pt
\begin{abstract}

\end{abstract}
 \vskip2pc]

\newpage

\medskip
{\bf Chizhov and Petcov Reply:} 
We have found in \cite{ChPet991}
{\it new conditions for a total 
neutrino conversion in a medium}
(see also \cite{ChPet992,SP98}).
It is claimed in \cite{ASPRL}
that our results are a particular case
of enhancement of neutrino oscillations, 
suggested in \cite{Param86,Akh88} 
and widely discussed 
in the literature.
We refute 
these claims,
confirming the 
novelty of our results.

  We have studied in \cite{ChPet991} 
the transitions
$\nu_{e} \rightarrow \nu_{\mu (\tau)}$, etc.
of neutrinos which 
crossed $n =2~(3)$ alternating 
layers with constant 
densities $N_{1}$ and $N_{2}$.
The probability of the transitions, 
$P_{2(3)} (\nu_a \rightarrow \nu_b)\equiv P_{2(3)}$,
is given by \cite{ChPet991,ChPet992}:
$$P_2 = 1 - Y^2 - X_{3}^2,~~~~~~
P_3 = 1 - \bar{Y}^2 - \bar{X}_{3}^2,~\eqno(1)$$
\noindent where $Y$, $X_3$ are defined
in \cite{ASPRL} (eqs. (1), (2)),
$\bar{Y} = - c_2 + 2c_1Y,~\bar{X}_{3} = 
- s_2 \cos 2\theta_2 - 2s_1 \cos (2\theta_1)Y$,
$c_{j}~(s_{j})\equiv \cos\phi_{j}~(\sin\phi_{j})$,
$\theta_{j}$ and 2$\phi_{j}$, $j=1,2$, 
being the mixing angle in matter in layer $j$
and the neutrino state 
phase difference
after neutrinos 
crossed this layer.
The {\it new} {\it conditions} 
{\it for 
a total neutrino conversion}
follow from (1) \cite{ChPet992}:
$$n=2:~~Y = 0,~X_3 = 0;~~~~
n=3:~~\bar{Y} = 0,~ \bar{X}_3 = 0.\eqno(2)$$
\noindent The solutions 
of these conditions were  given in 
\cite{ChPet991} (eqs. IV and (22)).
We have shown also \cite{ChPet991} 
that, e.g., for $n=2$, (2) are conditions for 
a {\it maximal constructive 
interference} between 
the amplitudes of neutrino 
transitions in the two layers.
Thus, a clear physical interpretation
of the absolute maxima of $P_2$ 
is that of {\it constructive 
interference maxima}.

(i) In connection with eq. (2) 
and the related effect 
of total neutrino conversion
the authors of \cite{ASPRL} write: 
``... the ``new effect of 
total neutrino conversion''
\cite{ChPet991} is nothing 
but a particular case
of the parametric resonance enhancement 
of neutrino oscillations, suggested in'' 
\cite{Param86,Akh88} ``and 
widely discussed in the literature...''.
We note
that the two sets of two  
conditions in (2) 
and their solutions 
were not derived and/or
discussed in any form  
in \cite{Param86,Akh88}
or in any other article 
published before \cite{ChPet991,ChPet992}. 
They do not follow from the conditions
of enhancement of 
$P(\nu_a \rightarrow \nu_b)$
found in \cite{Param86,Akh88} 
or \cite{KS89}.

(ii) For $n$ alternating layers 
one has according to 
\cite{ASPRL} 
$$ P_n(\nu_a \rightarrow \nu_b) = 
\frac{X_1^2 + X_2^2}{X_1^2 + X_2^2  + X_3^2}\sin^2 \Phi_p,~~\eqno(3)
$$
\noindent where ${\bf X} = (X_1,X_2,X_3)$ 
is a real vector, ${\bf X}^2 = 1 - Y^2$;
${\bf X}$ and $\Phi_p$ 
are defined in \cite{ASPRL} (eqs. (1) - (5)). 
According to \cite{ASPRL}, eq. (3) 
describes parametric oscillations,
and parametric resonance occurs when 
$X_3\equiv -(s_1c_2 \cos 2\theta_1+c_1s_2 \cos 2\theta_2)=0$, 
the latter being
the ``resonance condition''.
Due to eq. (1), 
any resonance interpretation  
of the probabilities
$P_{2,3}$
based solely on eq. (3) 
seems physically questionable.
Actually,
the denominator in (3) is 
always canceled by $\sin^2\Phi_p$, 
and $P_n$ 
is just  a polynomial without
any explicit resonance features: 
for even $n$, 
$\sin^2\Phi_p = {\bf X}^2U^{2}_{n/2-1}(Y)$,
$U_n(x)$ being the
well-known Chebyshev polynomial of the 
second kind, $Y = \cos\Phi =c_1c_2 - 
\cos(2\theta_2 - 2\theta_1)s_1s_2$;
for odd 
$n \geq 3$,
$$P_n=\left[
s_1\sin2\theta_1\cos\left(\frac{n-1}{2}\Phi\right)+
Z U_{\textstyle\frac{n-3}{2}}(Y)
\right]^2,\eqno(4)$$
\noindent 
$Z = s_2\sin 2\theta_2 + Y s_1\sin 2\theta_1$.
As should also be clear from (1), 
$X_{3} = 0$ 
{\it alone does not 
ensure the existence 
even of a local maximum} of 
$P_{2,3}$.
The same conclusion is valid 
for {\it any finite} $n$.

(iii) 
The two conditions $c_{1,2} = 0$
were not given in  
\cite{Param86,Akh88,KS89};
they were discussed first in \cite{SP98}.
What one finds in \cite{Param86,Akh88,KS89}
{\it at most} is 
$2\phi_1+2\phi_2=2\pi+2\pi k$. 
In addition, $c_{1,2} = 0$
are conditions of maxima of 
$P_{3,2}$ 
{\it only if fulfilled  
in a certain region
of the space of parameters}
\cite{SP98,ChPet991,ChPet992}.
Solution III is 
more than 
just $c_{1,2} = 0$ \cite{ChPet991}: 
{\it it includes as an 
integral part the region
$\cos(2\theta_2 - 2\theta_1) \leq 0$, 
where $c_{1,2} = 0$ have to hold}.
Thus, solution III in \cite{ChPet991} 
is not ``reproducing''
any of the solutions 
in \cite{Param86,Akh88,KS89}.
Moreover \cite{ChPet992},
$c_{1,2} = 0$
{\it lead to a maximum  
of $P_{2(3)}$
in the neutrino energy 
only when they hold
on the line}
$\cos(2\theta_2 - 2\theta_1) = 0$ 
($\cos(2\theta_2 - 4\theta_1) = 0$). 
The {\it three} conditions represent
a solution of (2) 
\cite{ChPet991,ChPet992}. 

(iv) {\it There is infinite  
number of irrelevant 
solutions of 
$X_3 =0$, $c_{1,2} \neq 0$, 
for $n=2,3$}. 
The existence 
of solution IV (or (22)),
its explicit form, found in \cite{ChPet991,ChPet992},
could not be, and were not, 
inferred from 
the solutions of 
$X_3 =0$, $c_{1,2} \neq 0$.

(v) 
The case of  neutrino oscillations 
in a medium with 3 layers, 
studied in  
\cite{SP98,ChPet991,ChPet992},
corresponds \cite{SP98} 
to the transitions 
in the Earth of the Earth-core-crossing
solar and atmospheric neutrinos.
Most importantly,
the {\it new conditions for
a total neutrino conversion} (2), 
found in \cite{ChPet991,ChPet992},
are fulfilled in this case 
and lead to observable effects 
\cite{ChPet991,ChPet992,SP98}.
The conditions of enhancement
of $P(\nu_a \rightarrow \nu_b)$
obtained in \cite{Param86,Akh88,KS89} 
are not valid
for the indicated 
$\nu-$transitions in the Earth. 

 Further comments on \cite{ASPRL} can be found in \cite{ChPet00}.


\end{document}